\documentclass{article}
\usepackage{ismir,amsmath,cite,url}
\usepackage{graphicx}
\usepackage{color}
\usepackage{booktabs}
\usepackage{multirow,subfigure}
\usepackage{float}

\title{Representation Learning of Music Using Artist Labels}

\multauthor
{Jiyoung Park$^1$$^*$ \hspace{0.1cm} Jongpil Lee$^2$$^*$ \hspace{0.1cm} Jangyeon Park$^1$ \hspace{0.1cm} Jung-Woo Ha$^1$ \hspace{0.1cm} Juhan Nam$^2$} {
$^1$ NAVER Corp.\\
$^2$ Graduate School of Culture Technology, KAIST\\
\footnote{$^*$ Equal Contribution}
{\tt\small \{j.y.park, jangyeon.park, jungwoo.ha\}@navercorp.com, \{richter, juhannam\}@kaist.ac.kr}
}
\def\authorname{Jiyoung Park, Jongpil Lee, Jangyeon Park, Jung-Woo Ha, Juhan Nam}

\begin{document}

\maketitle
\begin{abstract}

In music domain, feature learning has been conducted mainly in two ways: unsupervised learning based on sparse representations or supervised learning by semantic labels such as music genre. However, finding discriminative features in an unsupervised way is challenging and supervised feature learning using semantic labels may involve noisy or expensive annotation. In this paper, we present a supervised feature learning approach using artist labels annotated in every single track as objective meta data. We propose two deep convolutional neural networks (DCNN) to learn the deep artist features. One is a plain DCNN trained with the whole artist labels simultaneously, and the other is a Siamese DCNN trained with a subset of the artist labels based on the artist identity. We apply the trained models to music classification and retrieval tasks in transfer learning settings. The results show that our approach is comparable to previous state-of-the-art methods, indicating that the proposed approach captures general music audio features as much as the models learned with semantic labels. Also, we discuss the advantages and disadvantages of the two models.



\end{abstract}
\vspace{-2mm}
\section{Introduction}\label{sec:introduction}

\makeatletter
\def\blfootnote{\xdef\@thefnmark{}\@footnotetext}
\makeatother
\blfootnote{* Equally contributing authors.}

Representation learning or feature learning has been actively explored in recent years as an alternative to feature engineering  \cite{bengio2013representation}. The data-driven approach, particularly using deep neural networks, has been applied to the area of music information retrieval (MIR) as well \cite{Humphrey2013}. In this paper, we propose a novel audio feature learning method using deep convolutional neural networks and artist labels.  

Early feature learning approaches are mainly based on unsupervised learning algorithms. Lee et al. used convolutional deep belief network to learn structured acoustic patterns from spectrogram \cite{lee2009unsupervised}. They showed that the learned features achieve higher performance than Mel-Frequency Cepstral Coefficients (MFCC) in genre and artist classification. Since then, researchers have applied various unsupervised learning algorithms such as sparse coding \cite{M.Henaff:11, J.Nam:12,C.Yeh:13,vaizman2014codebook}, K-means \cite{J.Nam:12, J.Wulfing:12, dieleman2013multiscale} and restricted Boltzmann machine \cite{Schlueter:11, J.Nam:12}. Most of them focused on learning a meaningful dictionary on spectrogram by exploiting sparsity. While these unsupervised learning approaches are promising in that it can exploit abundant unlabeled audio data, most of them are limited to single or dual layers, which are not sufficient to represent complicated feature hierarchy in music. 



On the other hand, supervised feature learning has been progressively more explored. An early approach was mapping a single frame of spectrogram to genre or mood labels via pre-trained deep neural networks and using the hidden-unit activations as audio features \cite{P.Hamel:10,E.Schmidt:11}. More recently, this approach was handled in the context of transfer learning using deep convolutional neural networks (DCNN) \cite{choi2017transfer,LeeNam2017}. Leveraging large-scaled datasets and recent advances in deep learning, they showed that the hierarchically learned features can be effective for diverse music classification tasks. However, the semantic labels that they use such as genre, mood or other timbre descriptions tend to be noisy as they are sometimes ambiguous to annotate or tagged from the crowd. Also, high-quality annotation by music experts is known to be highly time-consuming and expensive. 

Meanwhile, artist labels are the meta data annotated to songs naturally from the album release. They are objective information with no disagreement. Furthermore, considering every artist has his/her own style of music, artist labels may be regarded as terms that describe diverse styles of music. Thus, if we have a model that can discriminate different artists from music, the model can be assumed to explain various characteristics of the music. 

\begin{figure*}[t!]
  \centering
  \subfigure[The Basic Model]{\includegraphics[height=6.1cm,keepaspectratio]{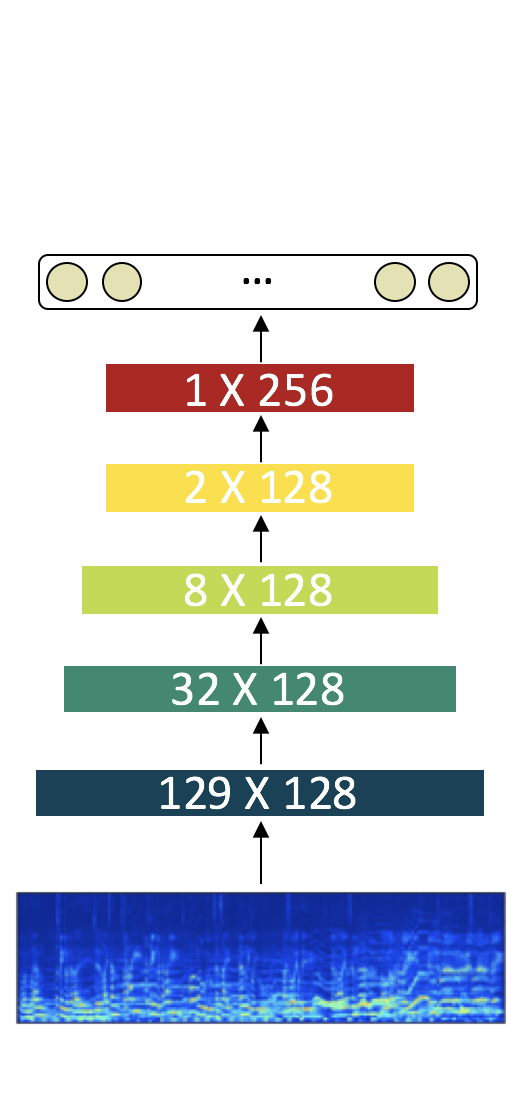}\label{fig:1a}}
  \subfigure[The Siamese Model]{\includegraphics[height=6.2cm,keepaspectratio]{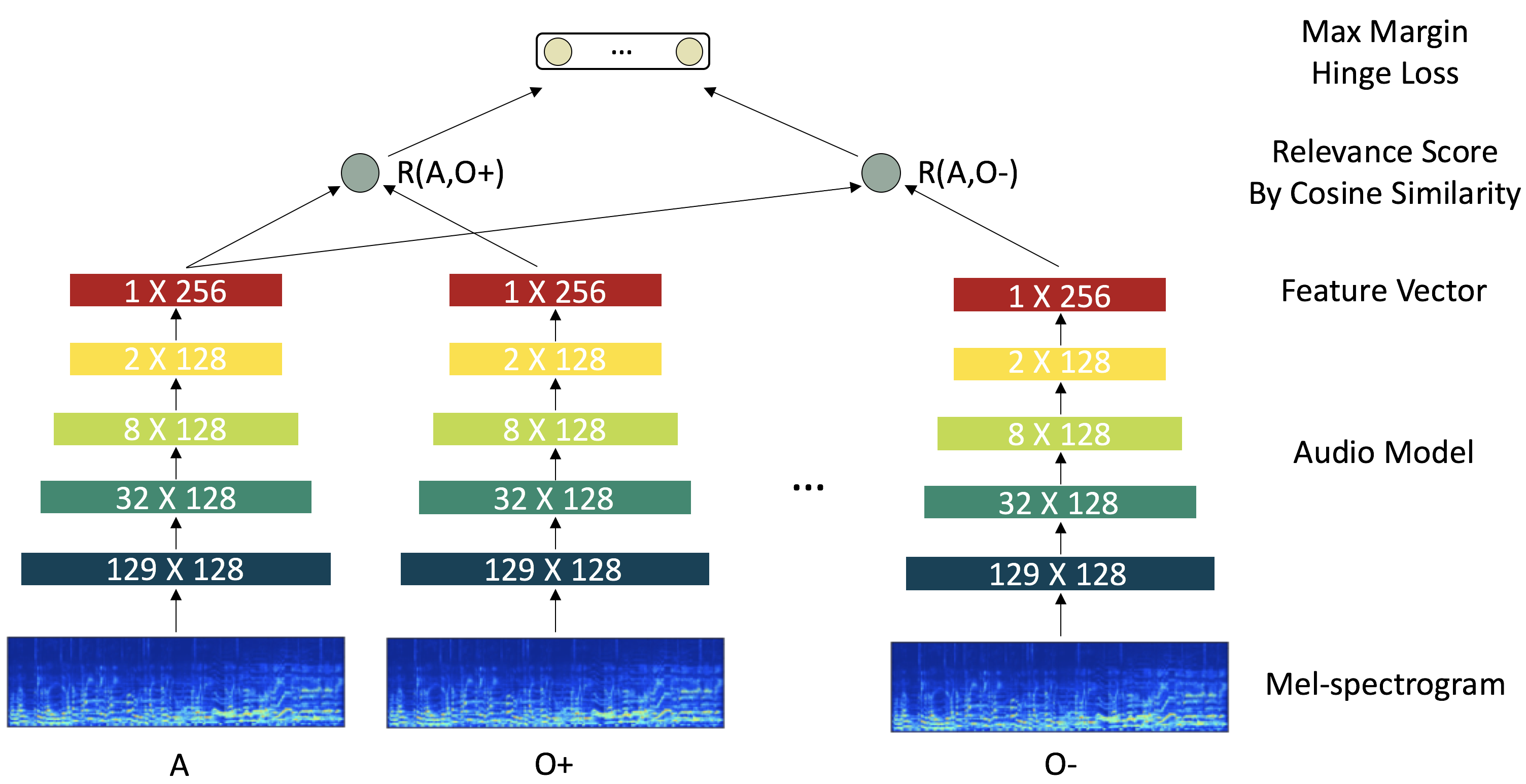}\label{fig:1b}}
  \vspace{-1.5mm}
  \caption{The proposed architectures for the model using artist labels.}
  \label{fig:fig1}
\end{figure*}

In this paper, we verify the hypothesis using two DCNN models that are trained to identify the artist from an audio track. One is the basic DCNN model where the softmax output units corresponds to each of artist. The other is the Siamese DCNN trained with a subset of the artist labels to mitigate the excessive size of the output layer in the plain DCNN when a large-scale dataset is used.  After training the two models, we regard them as a feature extractor and apply artist features to three different genre datasets in two experiment settings. First, we directly find similar songs using the artist features and K-nearest neighbors. Second, we conduct transfer learning to further adapter the features to each of the datasets. The results show that proposed approach captures useful features for unseen audio datasets and the propose models are comparable to those trained with semantic labels in performance. In addition, we discuss the advantages and disadvantages of the two proposed DCNN models.



\vspace{-2mm}
\section{Learning Models}
Figure \ref{fig:fig1} shows the two proposed DCNN models to learn audio features using artist labels. The basic model is trained as a standard classification problem. The Siamese model is trained using pair-wise similarity between an anchor artist and other artists. In this section, we describe them in detail.


\vspace{-1.5mm}

\subsection{Basic Model}


This is a widely used 1D-CNN model for music classification\cite{dieleman2014end,LeeNam2017,choi2017convolutional,pons2016experimenting}. The model uses mel-spectrogram with 128 bins in the input layer. We configured the DCNN such that one-dimensional convolution layers slide over only a single temporal dimension. The model is composed of 5 convolution and max pooling layers as illustrated in Figure \ref{fig:1a}. Batch normalization \cite{ioffe2015batch} and rectified linear unit (ReLU) activation layer are used after every convolution layer. Finally, we used categorical cross entropy loss in the prediction layer. 


We train the model to classify artists instead of semantic labels used in many music classification tasks. For example, if the number of artists used is 1,000, this becomes a classification problem that identifies one of the 1,000 artists. After training, the extracted 256-dimensional feature vector in the last hidden layer is used as the final audio feature learned using artist labels. Since this is the representation from which the identity is predicted by the linear softmax classifier, we can regard it as the highest-level artist feature.


\vspace{-1.5mm}

\subsection{Siamese Model}
While the basic model is simple to train, it has two main limitations. One is that the output layer can be excessively large if the dataset has numerous artists. For example, if a dataset has 10,000 artists and the last hidden layer size is 100, the number of parameters to learn in the last weight matrix will reach 1M. Second, whenever new artists are added to the dataset, the model must be trained again entirely. We solve the limitations using the Siamese DCNN model.  


A Siamese neural network consists of twin networks that share weights and configuration. It then provides unique inputs to the network and optimizes similarity scores \cite{bromley1994signature,koch2015siamese,manocha2017content}. This architecture can be extended to use both positive and negative examples at one optimization step. It is set up to take three examples: anchor item (query song), positive item (relevant song to the query) and negative item (different song to the query). This model is often called \emph{triplet networks} and has been successfully applied to music metric learning when the relative similarity scores of song triplets are available \cite{lu2017deep}. This model can be further extended to use several negative samples instead of just one negative in the triplet network. This technique is called \emph{negative sampling} and has been popularly used in word embedding \cite{mikolov2013distributed} and latent semantic model \cite{huang2013learning}. By using this technique, they could effectively approximate the full softmax function when the output class is extremely large (i.e. 10,000 classes).

We approximate the full softmax output in the basic model with the Siamese neural networks using negative sampling technique. Regarding the artist labels, we set up the negative sampling by treating identical artist's song to the anchor song as positive sample and other artists' songs as negative samples.
This method is illustrated in Figure \ref{fig:1b}. Following \cite{huang2013learning}, the relevance score between the anchor song feature and other song feature is measured as: 
\vspace{-2mm}
\begin{equation} \label{eq:cosine}
R(A,O) = \cos(y_A,y_O) = \frac{y_A^Ty_O}{|y_A||y_O|}
\end{equation}
where $y_A$ and $y_O$ are the feature vectors of the anchor song and other song, respectively. 

Meanwhile, the choice of loss function is important in this setting. We tested two loss functions. One is the softmax function with categorical cross-entropy loss to maximize the positive relationships. The other is the max-margin hinge loss to set only margins between positive and negative examples \cite{frome2013devise}. In our preliminary experiments, the Siamese model with negative sampling was successfully trained only with the max-margin loss function between the two objectives, which is defined as follows:

\vspace{-2mm}
\begin{equation} \label{eq:cosine}
loss(A,O) = \sum\limits_{O^-}\max[0, \Delta - R(A,O^+) + R(A,O^-)]
\end{equation}
where $\Delta$ is the margin, $O^+$ and $O^-$ denotes positive example and negative examples, respectively. We also grid-searched the number of negative samples and the margin, and finally set the number of negative samples to 4 and the margin value $\Delta$ to 0.4. The shared audio model used in this approach is exactly the same configuration as the basic model. 

\vspace{-2mm}
\subsection{Compared Model}
In order to verify the usefulness of the artist labels and the presented models, we constructed another model that has the same architecture as the basic model but using semantic tags. In this model, the output layer size corresponds to the number of the tag labels. Hereafter, we categorize all of them into \emph{artist-label model} and \emph{tag-label model}, and compare the performance. 


\vspace{-2mm}
\section{Experiments}

In this section, we describe source datasets to train the two artist-label models and one tag-label model. We also introduce target datasets for evaluating the three models. Finally, the training details are explained. 

\subsection{Source Tasks}

All models are trained with the Million Song Dataset (MSD)\cite{bertin2011million} along with 30-second 7digital\footnote{https://www.7digital.com/} preview clips. Artist labels are naturally annotated onto every song, thus we simply used them. For the tag label, we used the Last.fm dataset augmented on MSD. This dataset contains tag annotation that matches the ID of the MSD.

\subsubsection{Artist-label Model}
The number of songs that belongs to each artist may be extremely skewed and this can make fair comparison among the three models difficult. Thus, we selected 20 songs for each artist evenly and filtered out the artists who have less than this. Also, we configured several sets of the artist lists to see the effect of the number of artists on the model performances (500, 1,000, 2,000, 5,000 and 10,000 artists). We then divided them into 15, 3 and 2 songs for training, validation and testing, respectively for the sets contain less than 10,000 artists. For the 10,000 artist sets, we partitioned them in 17, 1 and 2 songs because once the artists reach 10,000, the validation set already become 10,000 songs even when we only use 1 song from each artist which is already sufficient for validating the model performance. We also should note that the testing set is actually not used in the whole experiments in this paper because we used the source dataset only for training the models to use them as feature extractors. The reason we filtered and split the data in this way is for future work\footnote{All the data splits of the source tasks are available at the link for reproducible research \url{https://github.com/jiyoungpark527/msd-artist-split}.}.



\subsubsection{Tag-label Model}

We used 5,000 artists set as a baseline experiment setting. This contains total 90,000 songs in the training and validation set with a split of 75,000 and 15,000. We thus constructed the same size set for tagging dataset to compare the artist-label models and the tag-label model. The tags and songs are first filtered in the same way as the previous works \cite{choi2016automatic,LeeNam2017}. Among the list with the filtered top 50 used tags, we randomly selected 90,000 songs and split them into the same size as the 5,000 artist set.



\subsection{Target Tasks}
We used 3 different datasets for genre classification.

\begin{itemize}
\item GTZAN (fault-filtered version) \cite{tzanetakis2002musical,kereliuk2015deep}: 930 songs, 10 genres. We used a ``fault-filtered'' version of GTZAN \cite{kereliuk2015deep} where the dataset was divided to prevent artist repetition in training/validation/test sets.
\item FMA small \cite{fma_dataset}: 8,000 songs, 8 balanced genres.
\item NAVER Music\footnote{http://music.naver.com} dataset with only Korean artists: 8,000 songs, 8 balanced genres. We filtered songs with only have one genre to clarify the genre characteristic.
\end{itemize}

\vspace{-2mm}
\subsection{Training Details} 

For the preprocessing, we computed the spectrogram using 1024 samples for FFT with a Hanning window, 512 samples for hop size and 22050 Hz as sampling rate. We then converted it to mel-spectrogram with 128 bins along with a log magnitude compression.  

We chose 3 seconds as a context window of the DCNN input after a set of experiments to find an optimal length that works well in music classification task. Out of the 30-second long audio, we randomly extracted the context size audio and put them into the networks as a single example. The input normalization was performed by dividing standard deviation after subtracting mean value across the training data.

We optimized the loss using stochastic gradient descent with 0.9 Nesterov momentum with $1e^{-6}$ learning rate decay. Dropout 0.5 is applied to the output of the last activation layer for all the models. We reduce the learning rate when a valid loss has stopped decreasing with the initial learning rate 0.015 for the basic models (both artist-label and tag-label) and 0.1 for the Siamese model. Zero-padding is applied to each convolution layer to maintain its size. 


Our system was implemented in Python 2.7, Keras 2.1.1 and Tensorflow-gpu 1.4.0 for the back-end of Keras. We used NVIDIA Tesla M40 GPU machines for training our models. Code and models are available at the link for reproducible research\footnote{\url{https://github.com/jongpillee/ismir2018-artist}.}.

\vspace{-2mm}
\section{Feature Evaluation}

We apply the learned audio features to genre classification as a target task in two different approaches: feature similarity-based retrieval and transfer learning. In this section, we describe feature extraction and feature evaluation methods.
\vspace{-2mm}
\subsection{Feature Extraction Using the DCNN Models}


In this work, the models are evaluated in three song-level genre classification tasks. Thus, we divided 30-second audio clip into 10 segments to match up with the model input size and the 256-dimension features from the last hidden layer are averaged into a single song-level feature vector and used for the following tasks. For the tasks that require song-to-song distances, cosine similarity is used to match up with the Siamese model's relevance score.

\vspace{-2mm}
\subsection{Feature Similarity-based Song Retrieval}

We first evaluated the models using mean average precision (MAP) considering genre labels as relevant items. After obtaining a ranked list for each song based on cosine similarity, we measured the MAP as following:
\begin{equation} \label{eq:AP}
AP = \frac{\sum_{k\in rel}precision_{k}}{number\ of\ relevant\ items}
\end{equation}
\begin{equation}
MAP = \frac{\sum_{q=1}^{Q}AP(q)}{Q}
\end{equation} 
where Q is the number of queries. \textit{$precision_k$} measures the fraction of correct items among first \textit{k} retrieved list.

The purpose of this experiment is to directly verify how similar feature vectors with the same genre are in the learned feature space.

\vspace{-2mm}
\subsection{Transfer Learning}

We classified audio examples using the k-nearest neighbors (k-NN) classifier and linear softmax classifier. The evaluation metric for this experiment is classification accuracy. We first classified audio examples using k-NN to classify the input audio into the largest number of genres among k nearest to features from the training set.
The number of k is set to 20 in this experiment. This method can be regarded as a similarity-based classification. 
We also classified audio using a linear softmax classifier. The purpose of this experiment is to verify how much the audio features of unseen datasets are linearly separable in the learned feature space. 

\begin{table}[!t]
\small
\centering
\resizebox{\columnwidth}{!}{\begin{tabular}{cccc}
\toprule
MAP          & \begin{tabular}[c]{@{}c@{}}Artist-label\\Basic Model\end{tabular} & \begin{tabular}[c]{@{}c@{}}Artist-label\\Siamese Model\end{tabular} & \begin{tabular}[c]{@{}c@{}}Tag-label\\Model\end{tabular} \\ \midrule
\begin{tabular}[c]{@{}c@{}}GTZAN\\(fault-filtered)\end{tabular}        & 0.4968          & 0.5510             & 0.5508               \\
FMA small    & 0.2441          & 0.3203             & 0.3019               \\
NAVER Korean & 0.3152          & 0.3577             & 0.3576             \\ \bottomrule
\end{tabular}}
\caption{MAP results on feature similarity-based retrieval.}
\label{MAP}
\end{table}

\begin{table}[!t]
\small
\centering
\resizebox{\columnwidth}{!}{\begin{tabular}{@{}cccc@{}}
\toprule
KNN          & \begin{tabular}[c]{@{}c@{}}Artist-label\\Basic Model\end{tabular} & \begin{tabular}[c]{@{}c@{}}Artist-label\\Siamese Model\end{tabular} & \begin{tabular}[c]{@{}c@{}}Tag-label\\Model\end{tabular} \\ \midrule
\begin{tabular}[c]{@{}c@{}}GTZAN\\(fault-filtered)\end{tabular}        & 0.6655          & 0.6966             & 0.6759               \\
FMA small    & 0.5269          & 0.5732             & 0.5332               \\
NAVER Korean & 0.6671          & 0.6393             & 0.6898               \\  \bottomrule
\end{tabular}}
\caption{KNN similarity-based classification accuracy.}
\label{KNN}
\end{table}

\begin{table}[!t]
\small
\centering
\resizebox{\columnwidth}{!}{\begin{tabular}{cccc}
\toprule
Linear Softmax          & \begin{tabular}[c]{@{}c@{}}Artist-label\\Basic Model\end{tabular} & \begin{tabular}[c]{@{}c@{}}Artist-label\\Siamese Model\end{tabular} & \begin{tabular}[c]{@{}c@{}}Tag-label\\Model\end{tabular} \\ \midrule
\begin{tabular}[c]{@{}c@{}}GTZAN\\(fault-filtered)\end{tabular}        & 0.6721          & 0.6993             & 0.7072               \\
FMA small    & 0.5791          & 0.5483             & 0.5641               \\
NAVER Korean & 0.6696          & 0.6623             & 0.6755               \\ \bottomrule
\end{tabular}}
\caption{Classification accuracy of a linear softmax.}
\label{softmax}
\end{table}

\begin{figure*}[!th]
\vspace{-1mm}
\small
 \centerline{{
 \includegraphics[width=2.1\columnwidth]{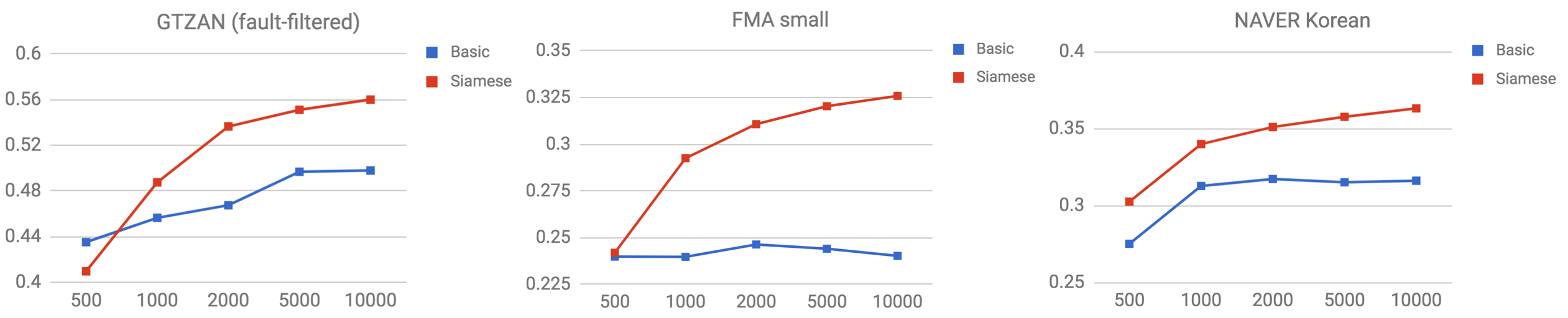}}}
 \vspace{-1.3mm}
 \caption{MAP results with regard to different number of artists in the feature models.}
\label{fig:numartmap}
\end{figure*}

\begin{figure*}[!th]
\vspace{-1mm}
\small
 \centerline{{
 \includegraphics[width=2.1\columnwidth]{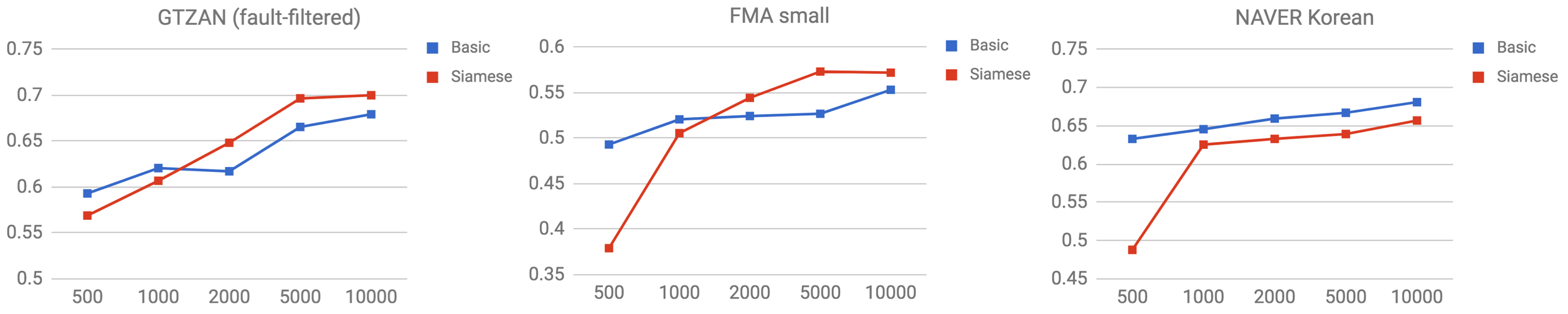}}}
 \vspace{-1.3mm}
 \caption{Genre classification accuracy using k-NN with regard to different number of artists in the feature models.}
\label{fig:numartknn}
\end{figure*}

\begin{figure*}[!th]
\vspace{-1mm}
\small
 \centerline{{
 \includegraphics[width=2.1\columnwidth]{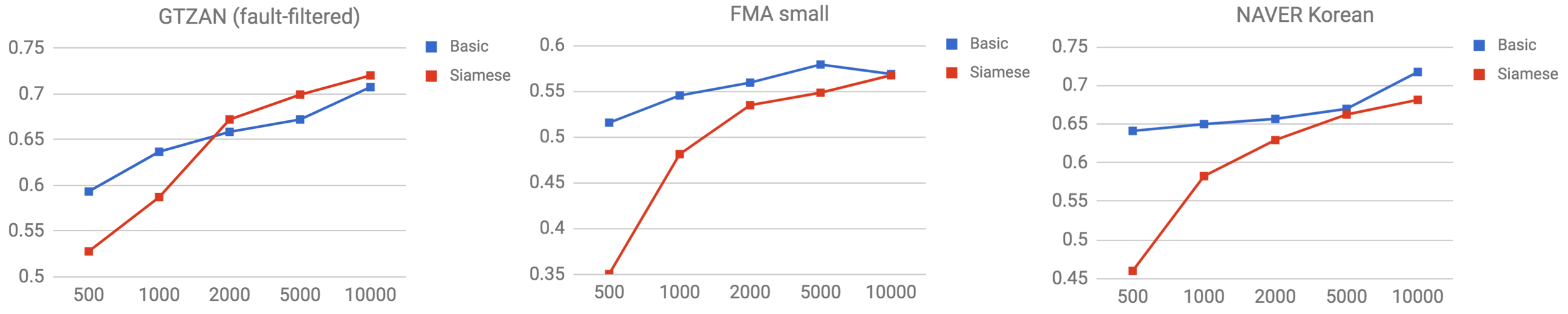}}}
 \vspace{-1.3mm}
 \caption{Genre classification accuracy using linear softmax with regard to different number of artists in the feature models.}
\label{fig:numartsoftmax}
\end{figure*}

\section{Results and Discussion}

\subsection{Tag-label Model vs. Artist-label Model}
We first compare the artist-label models to the tag-label model when they are trained with the same dataset size (90,000 songs). The results are shown in Table \ref{MAP}, \ref{KNN} and \ref{softmax}. In feature similarity-based retrieval using MAP (Table \ref{MAP}), the artist-based Siamese model outperforms the rest on all target datasets. In the genre classification tasks (Table {\ref{KNN} and \ref{softmax}), Tag-label model works slightly better than the rest on some datasets and the trend becomes stronger in the classification using the linear softmax. Considering that the source task in the tag-based model (trained with the Last.fm tags) contains genre labels mainly, this result may attribute to the similarity of labels in both source and target tasks. Therefore, we can draw two conclusions from this experiment. First, the artist-label model is more effective in similarity-based tasks (\ref{MAP} and \ref{KNN}) when it is trained with the proposed Siamese networks, and thus it may be more useful for music retrieval. Second, the semantic-based model is more effective in genre or other semantic label tasks and thus it may be more useful for human-friendly music content organization. 




\subsection{Basic Model vs. Siamese Model}
\label{basic_siamese}
Now we focus on the comparison of the two artist-label models. From Table \ref{MAP}, \ref{KNN} and \ref{softmax}, we can see that the Siamese model generally outperforms the basic model. However, the difference become attenuated in classification tasks and the Siamese model is even worse on some datasets. 
Among them, it is notable that the Siamese model is significantly worse than the basic model on the NAVER Music dataset in the genre classification using k-NN even though they are based on feature similarity. We dissected the result to see whether it is related to the cultural difference between the training data (MSD, mostly Western) and the target data (the NAVER set, only Korean). Figure \ref{fig:korean_genre} shows the detailed classification accuracy for each genre of the NAVER dataset. In three genres, `Trot',`K-pop Ballad' and `Kids' that do not exist in the training dataset, we can see that the basic model outperforms the Siamese model whereas the results are opposite in the other genres. This indicates that the basic model is more robust to unseen genres of music. On the other hand, the Siamese model slightly over-fits to the training set, although it effectively captures the artist features. 




\vspace{-3mm}
\begin{figure}[t]
\small
 \centerline{{
 \includegraphics[keepaspectratio,height=4.5cm]{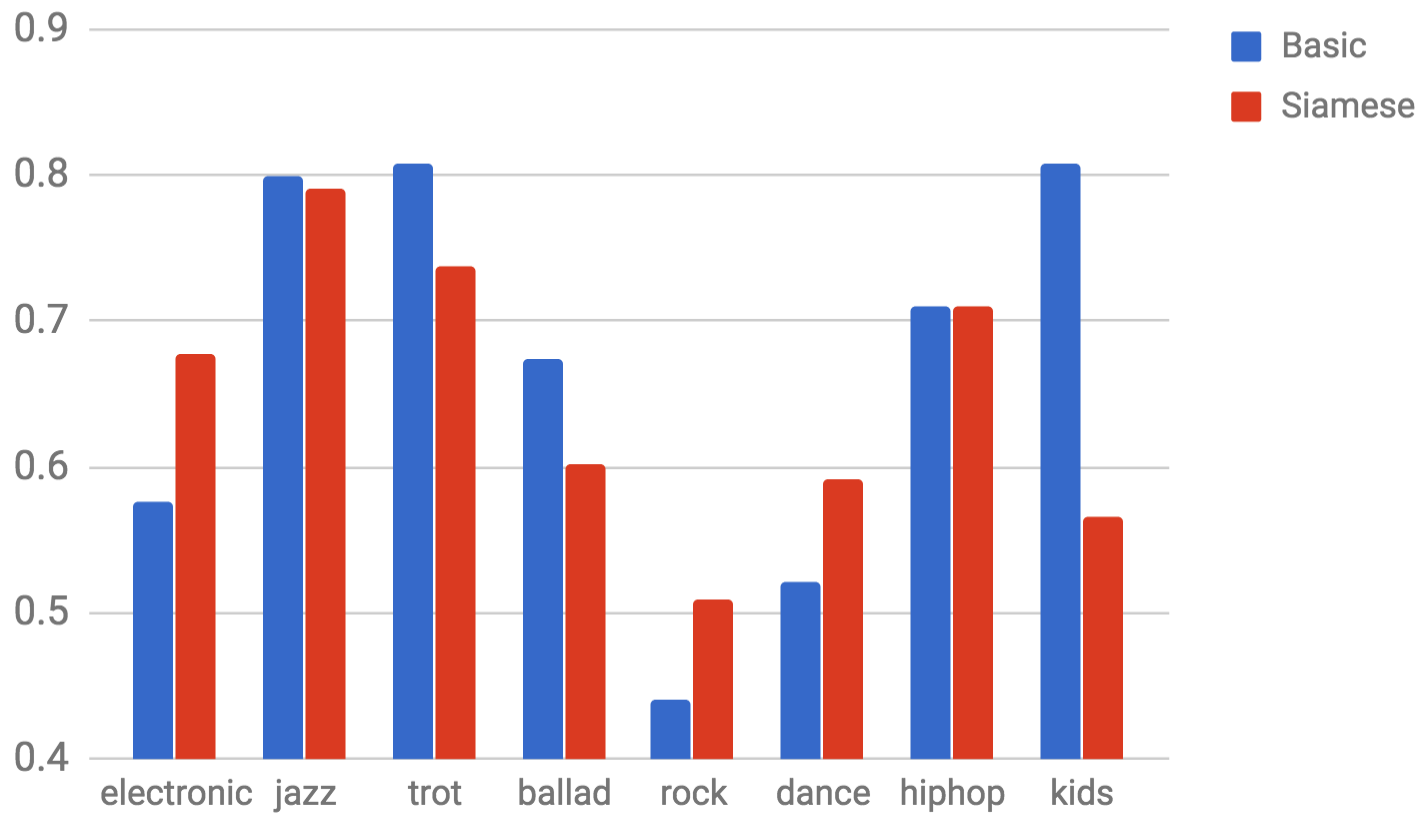}}}
 \vspace{-3mm}
 \caption{The classification results of each genre for the NAVER dataset with only Korean music.}
\label{fig:korean_genre}
\end{figure}



\begin{table}[!t]
\small
\centering
\resizebox{\columnwidth}{!}{\begin{tabular}{lccc}
\hline \hline
Models                  & \begin{tabular}[c]{@{}c@{}}GTZAN\\(fault-filtered)\end{tabular}  & FMA small \\ \hline
2-D CNN \cite{kereliuk2015deep}                & 0.6320  & -   \\
Temporal features \cite{jeong2016learning}      & 0.6590  & -   \\
Multi-level Multi-scale \cite{LeeNam2017} & 0.7200 & - \\ 
SVM \cite{fma_dataset} &-  & 0.5482$^{\dagger}$ \\ \hline
Artist-label Basic model   & 0.7076 &  0.5687  \\ 
Artist-label Siamese model   &0.7203  &  0.5673  \\ \hline \hline
\end{tabular}}
\caption{Comparison with previous state-of-the-art models: classification accuracy results. Linear softmax classifier is used and features are extracted from the artist-label models trained with 10,000 artists. $\dagger$ This result was obtained using the provided code and dataset in \cite{fma_dataset}.}
\label{tab:state_arts}
\end{table}

\subsection{Effect of the Number of Artists} 

We further analyze the artist-label models by investigating how the number of artists in training the DCNN affects the performance. Figure \ref{fig:numartmap}, \ref{fig:numartknn} and \ref{fig:numartsoftmax} are the results that show similarity-based retrieval (MAP) and genre classification (accuracy) using k-NN and linear softmax, respectively, according to the increasing number of training artists. They show that the performance is generally proportional to the number of artists but the trends are quite different between the two models. In the similarity-based retrieval, the MAP of the Siamese model is significantly higher than that of the basic model when the number of artists is greater than 1,000. Also, as the number of artists increases, the MAP of the Siamese model consistently goes up with a slight lower speed whereas that of the basic model saturates at 2,000 or 5,000 artists. On the other hand, the performance gap changes in the two classification tasks. On the GTZAN dataset, while the basic model is better for 500 and 1,000 artists, the Siamese model reverses it for 2,000 and more artists. On the NAVER dataset, the basic model is consistently better. On the FMA small, the results are mixed in two classifiers. Again, the results may be explained by our interpretation of the models in Section \ref{basic_siamese}. In summary, the Siamese model seems to work better in similarity-based tasks and the basic model is more robust to different genres of music. In addition, the Siamese model is more capable of being trained with a large number of artists. 

\vspace{-2mm}
\subsection{Comparison with State-of-the-arts}
The effectiveness of artist labels is also supported by comparison with previous state-of-the-art models in Table \ref{tab:state_arts}. For this result, we report two artist-label models trained with 10,000 artists using linear softmax classifier. 
In this table, we can see that the proposed models are comparable to the previous state-of-the-art methods. 


\vspace{-2.5mm}
\section{Visualization}
We visualize the extracted feature to provide better insight on the discriminative power of learned features using artist labels. We used the DCNN trained to classify 5,000 artists as a feature extractor. After collecting the feature vectors, we embedded them into 2-dimensional vectors using t-distributed stochastic neighbor embedding (t-SNE). 

For artist visualization, we collect a subset of MSD (apart from the training data for the DCNN) from well-known artists. Figure \ref{fig:artist_viz} shows that artists' songs are appropriately distributed based on genre, vocal style and gender. For example, artists with similar genre of music are closely located and female pop singers are close to each other except Maria Callas who is a classical opera singer. Interestingly, some songs by Michael Jackson are close to female vocals because of his distinctive high-pitched tone. 

Figure \ref{fig:genre_viz} shows the visualization of features extracted from the GTZAN dataset. Even though the DCNN was trained to discriminate artist labels, they are well clustered by genre. Also, we can observe that some genres such as disco, rock and hip-hop are divided into two or more groups that might belong to different sub-genres.

\begin{figure}[t]
\vspace{-3mm}
\small
 \centerline{{
 \includegraphics[width=8cm,height=5.3cm]{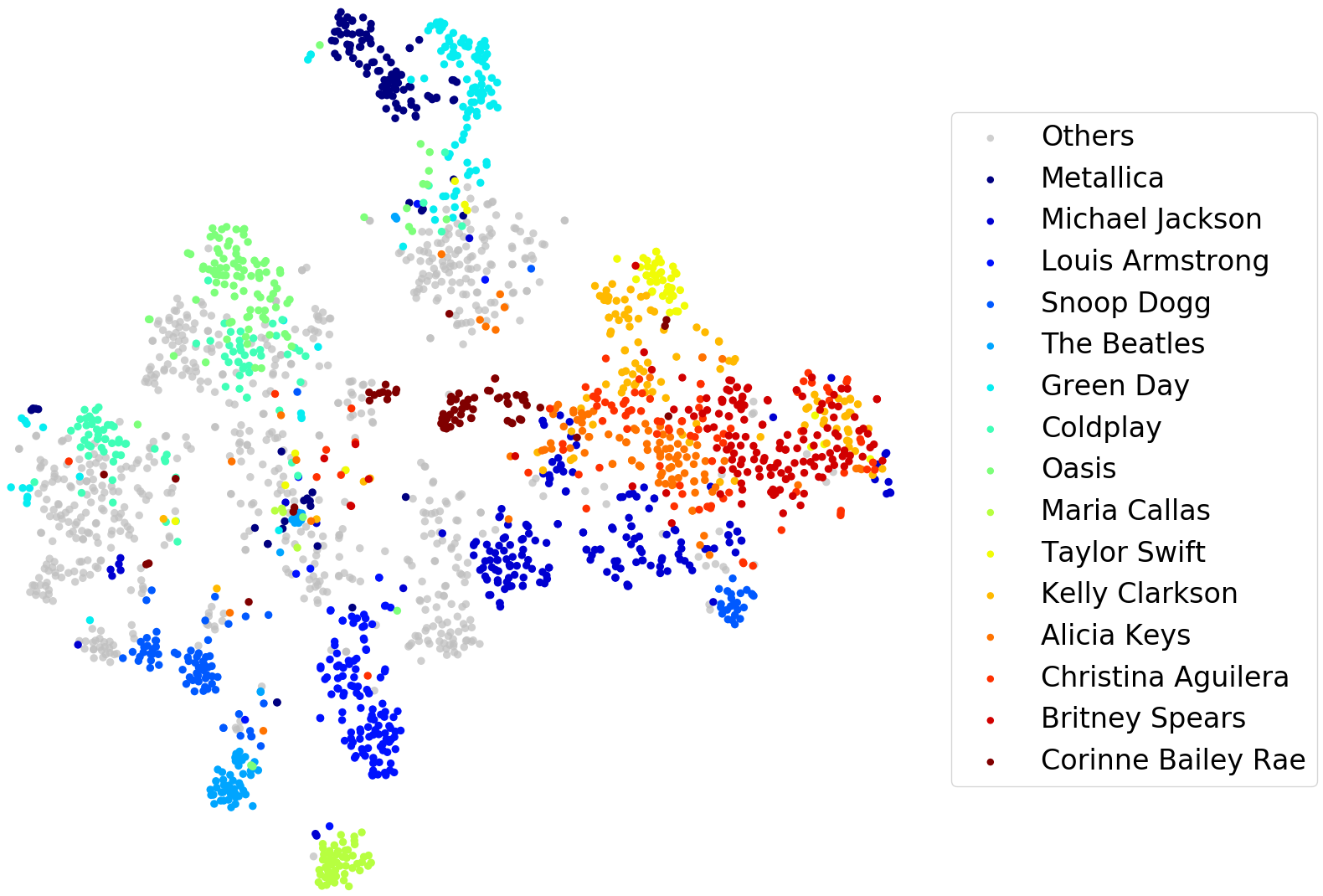}}}
 \caption{Feature visualization by artist. Total 22 artists are used and, among them, 15 artists are represented in color.}
\label{fig:artist_viz}
\end{figure}

\begin{figure}[t]
\vspace{-5mm}
\small
 \centerline{{
 \includegraphics[width=6.7cm,height=5cm]{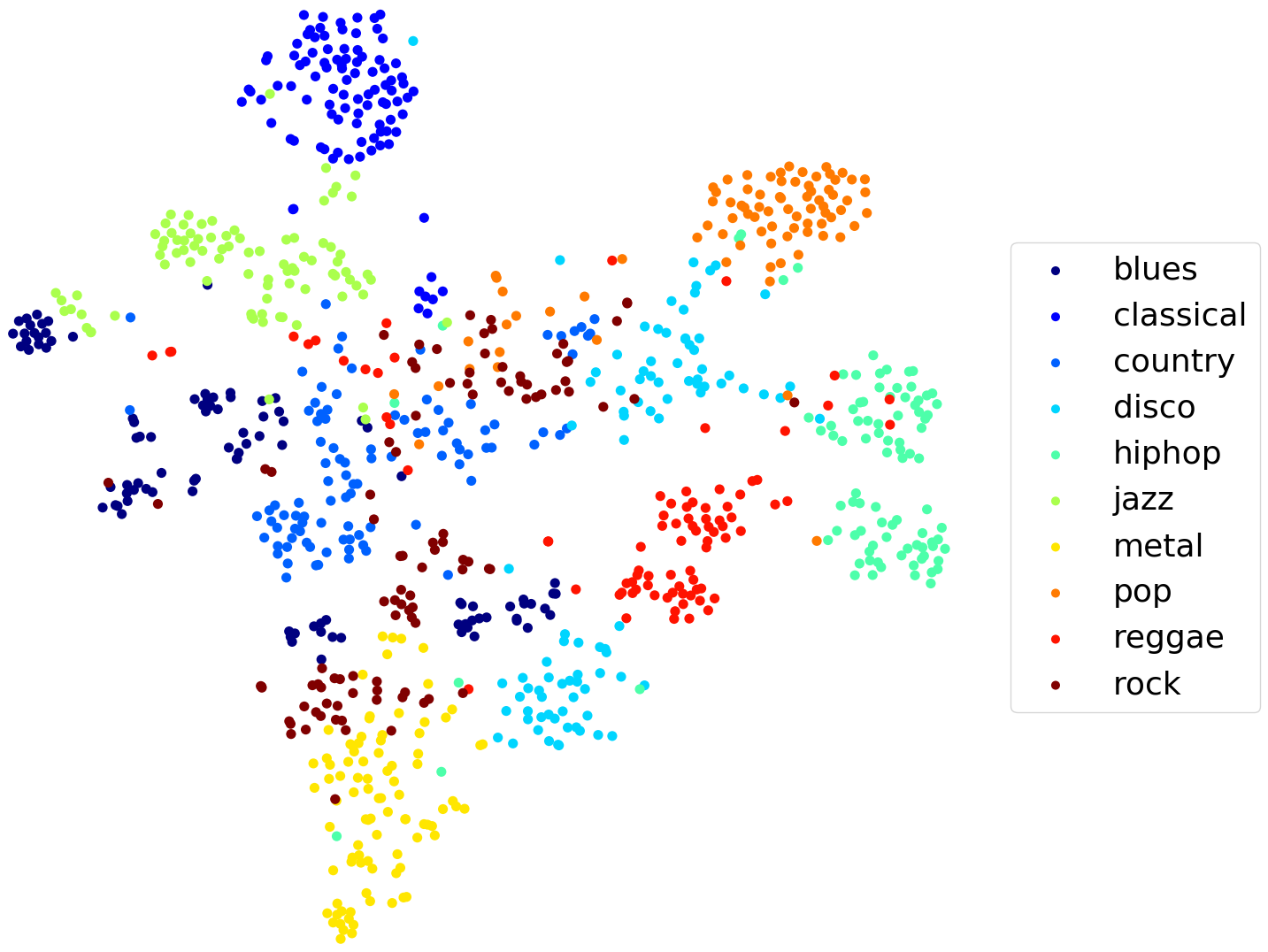}}}
 \caption{Feature visualization by genre. Total 10 genres from the GTZAN dataset are used.}
\label{fig:genre_viz}
\end{figure}

\vspace{-2.5mm}
\section{Conclusion and Future work}

In this work, we presented the models to learn audio feature representation using artist labels instead of semantic labels. We compared two artist-label models and one tag-label model. The first is a basic DCNN consisting of a softmax output layer to predict which artist they belong to out of all artists used. The second is a Siamese-style architecture that maximizes the relative similarity score between a small subset of the artist labels based on the artist identity. The last is a model optimized using tag labels with the same architecture as the first model. After the models are trained, we used them as feature extractors and validated the models on song retrieval and genre classification tasks on three different datasets. Three interesting results were found during the experiments. First, the artist-label models, particularly the Siamese model, is comparable to or outperform the tag-label model. This indicates that the cost-free artist-label is as effective as the expensive and possibly noisy tag-label. Second, the Siamese model showed the best performances on song retrieval task in all datasets tested. This can indicate that the pair-wise relevance score loss in the Siamese model helps the feature similarity-based search. Third, the use of a large number of artists increases the model performance. This result is also useful because the artists can be easily increased to a very large number. 

As future work, we will investigate the artist-label Siamese model more thoroughly. First, we plan to investigate advanced audio model architecture and diverse loss and pair-wise relevance score functions. Second, the model can easily be re-trained using new added artists because the model does not have fixed output layer. This property will be evaluated using cross-cultural data or using extremely small data (i.e. one-shot learning \cite{koch2015siamese}). 

\section{Acknowledgement}
This work was supported by Basic Science Research Program through the National Research Foundation of Korea funded by the Ministry of  Science, ICT \& Future Planning (2015R1C1A1A02036962) and by NAVER Corp.

\bibliography{ISMIRtemplate}

\begin{thebibliography}{10}

\bibitem{bengio2013representation}
Yoshua Bengio, Aaron Courville, and Pascal Vincent.
\newblock Representation learning: A review and new perspectives.
\newblock {\em IEEE transactions on pattern analysis and machine intelligence},
  35(8):1798--1828, 2013.

\bibitem{bertin2011million}
Thierry Bertin-Mahieux, Daniel~PW Ellis, Brian Whitman, and Paul Lamere.
\newblock The million song dataset.
\newblock In {\em Proc. of the International Society for Music Information
  Retrieval Conference (ISMIR)}, volume~2, pages 591--596, 2011.

\bibitem{bromley1994signature}
Jane Bromley, Isabelle Guyon, Yann LeCun, Eduard S{\"a}ckinger, and Roopak
  Shah.
\newblock Signature verification using a ``siamese'' time delay neural network.
\newblock In {\em Advances in Neural Information Processing Systems (NIPS)},
  pages 737--744, 1994.

\bibitem{choi2016automatic}
Keunwoo Choi, George Fazekas, and Mark Sandler.
\newblock Automatic tagging using deep convolutional neural networks.
\newblock In {\em Proc. of the International Society for Music Information
  Retrieval Conference (ISMIR)}, pages 805--811, 2016.

\bibitem{choi2017convolutional}
Keunwoo Choi, Gy{\"o}rgy Fazekas, Mark Sandler, and Kyunghyun Cho.
\newblock Convolutional recurrent neural networks for music classification.
\newblock In {\em Proc. of the IEEE International Conference on Acoustics,
  Speech, and Signal Processing (ICASSP)}, pages 2392--2396, 2017.

\bibitem{choi2017transfer}
Keunwoo Choi, Gy{\"o}rgy Fazekas, Mark Sandler, and Kyunghyun Cho.
\newblock Transfer learning for music classification and regression tasks.
\newblock In {\em Proc. of the International Conference on Music Information
  Retrieval (ISMIR)}, pages 141--149, 2017.

\bibitem{fma_dataset}
Micha\"el Defferrard, Kirell Benzi, Pierre Vandergheynst, and Xavier Bresson.
\newblock Fma: A dataset for music analysis.
\newblock In {\em Proc. of the International Society for Music Information
  Retrieval Conference (ISMIR)}, pages 316--323, 2017.

\bibitem{dieleman2013multiscale}
Sander Dieleman and Benjamin Schrauwen.
\newblock Multiscale approaches to music audio feature learning.
\newblock In {\em International Society for Music Information Retrieval
  Conference (ISMIR)}, pages 116--121, 2013.

\bibitem{dieleman2014end}
Sander Dieleman and Benjamin Schrauwen.
\newblock End-to-end learning for music audio.
\newblock In {\em Proc. of the IEEE International Conference on Acoustics,
  Speech and Signal Processing (ICASSP)}, pages 6964--6968, 2014.

\bibitem{frome2013devise}
Andrea Frome, Greg~S Corrado, Jon Shlens, Samy Bengio, Jeff Dean, Tomas
  Mikolov, et~al.
\newblock Devise: A deep visual-semantic embedding model.
\newblock In {\em Advances in neural information processing systems (NIPS)},
  pages 2121--2129, 2013.

\bibitem{P.Hamel:10}
Philippe Hamel and Douglas Eck.
\newblock Learning features from music audio with deep belief networks.
\newblock In {\em Proc. of the International Conference on Music Information
  Retrieval (ISMIR)}, pages 339--344, 2010.

\bibitem{M.Henaff:11}
Mikael Henaff, Kevin Jarrett, Koray Kavukcuoglu, and Yann LeCun.
\newblock Unsupervised learning of sparse features for scalable audio
  classification.
\newblock In {\em Proc. of the International Conference on Music Information
  Retrieval (ISMIR)}, pages 681--686, 2011.

\bibitem{huang2013learning}
Po-Sen Huang, Xiaodong He, Jianfeng Gao, Li~Deng, Alex Acero, and Larry Heck.
\newblock Learning deep structured semantic models for web search using
  clickthrough data.
\newblock In {\em Proc. of the 22nd ACM international conference on Conference
  on information \& knowledge management}, pages 2333--2338. ACM, 2013.

\bibitem{Humphrey2013}
Eric Humphrey, Juan Bello, and Yann LeCun.
\newblock Feature learning and deep architectures: new directions for music
  informatics.
\newblock {\em Journal of Intelligent Information Systems}, 41(3):461--481, Dec
  2013.

\bibitem{ioffe2015batch}
Sergey Ioffe and Christian Szegedy.
\newblock Batch normalization: Accelerating deep network training by reducing
  internal covariate shift.
\newblock In {\em International Conference on Machine Learning (ICML)}, pages
  448--456, 2015.

\bibitem{jeong2016learning}
Il-Young Jeong and Kyogu Lee.
\newblock Learning temporal features using a deep neural network and its
  application to music genre classification.
\newblock In {\em Proc. of the International Society for Music Information
  Retrieval Conference (ISMIR)}, pages 434--440, 2016.

\bibitem{kereliuk2015deep}
Corey Kereliuk, Bob~L Sturm, and Jan Larsen.
\newblock Deep learning and music adversaries.
\newblock {\em IEEE Transactions on Multimedia}, 17(11):2059--2071, 2015.

\bibitem{koch2015siamese}
Gregory Koch, Richard Zemel, and Ruslan Salakhutdinov.
\newblock Siamese neural networks for one-shot image recognition.
\newblock In {\em ICML Deep Learning Workshop}, volume~2, 2015.

\bibitem{lee2009unsupervised}
Honglak Lee, Peter Pham, Yan Largman, and Andrew~Y Ng.
\newblock Unsupervised feature learning for audio classification using
  convolutional deep belief networks.
\newblock In {\em Advances in neural information processing systems (NIPS)},
  pages 1096--1104, 2009.

\bibitem{LeeNam2017}
Jongpil Lee and Juhan Nam.
\newblock Multi-level and multi-scale feature aggregation using pretrained
  convolutional neural networks for music auto-tagging.
\newblock {\em IEEE Signal Processing Letters}, 24(8):1208--1212, 2017.

\bibitem{lu2017deep}
Rui Lu, Kailun Wu, Zhiyao Duan, and Changshui Zhang.
\newblock Deep ranking: Triplet matchnet for music metric learning.
\newblock In {\em Proc. of the IEEE International Conference on Acoustics,
  Speech and Signal Processing (ICASSP)}, pages 121--125, 2017.

\bibitem{manocha2017content}
Pranay Manocha, Rohan Badlani, Anurag Kumar, Ankit Shah, Benjamin Elizalde, and
  Bhiksha Raj.
\newblock Content-based representations of audio using siamese neural networks.
\newblock In {\em Proc. of the IEEE International Conference on Acoustics,
  Speech and Signal Processing (ICASSP)}, 2018.

\bibitem{mikolov2013distributed}
Tomas Mikolov, Ilya Sutskever, Kai Chen, Greg~S Corrado, and Jeff Dean.
\newblock Distributed representations of words and phrases and their
  compositionality.
\newblock In {\em Advances in neural information processing systems (NIPS)},
  pages 3111--3119, 2013.

\bibitem{J.Nam:12}
Juhan Nam, Jorge Herrera, Malcolm Slaney, and Julius~O. Smith.
\newblock Learning sparse feature representations for music annotation and
  retrieval.
\newblock In {\em Proc. of the International Conference on Music Information
  Retrieval (ISMIR)}, pages 565--570, 2012.

\bibitem{pons2016experimenting}
Jordi Pons, Thomas Lidy, and Xavier Serra.
\newblock Experimenting with musically motivated convolutional neural networks.
\newblock In {\em Proc. of the International Workshop on Content-Based
  Multimedia Indexing (CBMI)}, pages 1--6, 2016.

\bibitem{Schlueter:11}
Jan Schl{\"u}ter and Christian Osendorfer.
\newblock {Music Similarity Estimation with the Mean-Covariance Restricted
  Boltzmann Machine}.
\newblock In {\em Proc. of the International Conference on Machine Learning and
  Applications}, pages 118--123, 2011.

\bibitem{E.Schmidt:11}
Erik~M. Schmidt and Youngmoo~E. Kim.
\newblock Learning emotion-based acoustic features with deep belief networks.
\newblock In {\em Proc. of the IEEE Workshop on Applications of Signal
  Processing to Audio and Acoustics (WASPAA)}, pages 65--68, 2011.

\bibitem{tzanetakis2002musical}
George Tzanetakis and Perry Cook.
\newblock Musical genre classification of audio signals.
\newblock {\em IEEE Transactions on speech and audio processing},
  10(5):293--302, 2002.

\bibitem{vaizman2014codebook}
Yonatan Vaizman, Brian McFee, and Gert Lanckriet.
\newblock Codebook-based audio feature representation for music information
  retrieval.
\newblock {\em IEEE/ACM Transactions on Audio, Speech and Language Processing
  (TASLP)}, 22(10):1483--1493, 2014.

\bibitem{J.Wulfing:12}
Jan W{\"u}lfing and Martin Riedmiller.
\newblock Unsupervised learning of local features for music classification.
\newblock In {\em Proc. of the International Society for Music Information
  Retrieval Conference (ISMIR)}, pages 139--144, 2012.

\bibitem{C.Yeh:13}
Chin-Chia Yeh, Li~Su, and Yi-Hsuan Yang.
\newblock Dual-layer bag-of-frames model for music genre classification.
\newblock In {\em Proc. of the IEEE International Conference on Acoustics,
  Speech, and Signal Processing (ICASSP)}, pages 246--250, 2013.

\end{thebibliography}

%
%
%
%

\end{document}